# Time-lapse image classification using a diffractive neural network

Md Sadman Sakib Rahman[1,2,3]    mssr@ucla.edu
Aydogan Ozcan[1,2,3,*]            ozcan@ucla.edu

[1]Electrical and Computer Engineering Department, University of California, Los Angeles, CA, 90095, USA
[2]Bioengineering Department, University of California, Los Angeles, CA, 90095, USA
[3]California NanoSystems Institute (CNSI), University of California, Los Angeles, CA, 90095, USA
[*]Corresponding author: ozcan@ucla.edu

**Abstract**

Diffractive deep neural networks ($D^2$NNs) define an all-optical computing framework comprised of spatially engineered passive surfaces that collectively process optical input information by modulating the amplitude and/or the phase of the propagating light. Diffractive optical networks complete their computational tasks at the speed of light propagation through a thin diffractive volume, without any external computing power while exploiting the massive parallelism of optics. Diffractive networks were demonstrated to achieve all-optical classification of objects and perform universal linear transformations. Here we demonstrate, for the first time, a 'time-lapse' image classification scheme using a diffractive network, significantly advancing its classification accuracy and generalization performance on complex input objects by using the lateral movements of the input objects and/or the diffractive network, relative to each other. In a different context, such relative movements of the objects and/or the camera are routinely being used for image super-resolution applications; inspired by their success, we designed a time-lapse diffractive network to benefit from the complementary information content created by controlled or random lateral shifts. We numerically explored the design space and performance limits of time-lapse diffractive networks, revealing a blind testing accuracy of 62.03% on the optical classification of objects from the CIFAR-10 dataset. This constitutes the highest inference accuracy achieved so far using a single diffractive network on the CIFAR-10 dataset. Time-lapse diffractive networks will be broadly useful for the spatio-temporal analysis of input signals using all-optical processors.

# Introduction

Machine learning and artificial intelligence research has experienced rapid growth in the past two decades[1]. One of the core engines that has driven this growth is deep learning[2], permitting efficient and rapid training of deep artificial neural network models. The ability to train deep neural networks has revolutionized artificial intelligence, and electronics has been the undisputed platform of choice for implementing artificial neural networks. Specialized processing hardware such as Graphics Processing Units (GPUs) are widely used today for deep learning. However, these electronic processors are power-hungry and bulky, making researchers wary of the environmental impact of machine learning[3,4]. Therefore, there is strong interest in low-power and fast computing platforms for machine learning applications. Optical computing has been identified as a promising potential alternative for such purposes because of the large bandwidth, high speed, and massive parallelism of optics[5].

Diffractive deep neural networks ($D^2$NNs), also known as diffractive optical networks or diffractive networks, form a passive all-optical computing platform that exploits the diffraction of light waves to perform computation[6]. These diffractive networks are composed of several spatially-engineered surfaces, separated by free-space. The diffractive features/elements of a layer, also termed 'diffractive neurons', locally modulate the amplitude and/or the phase of the light incident upon the layer. Successive modulation by and diffraction through the layers give rise to an all-optical transformation between the input and the output fields-of-view at the speed of light propagation without any external power. The amplitude and/or the phase values of the diffractive neurons corresponding to a desired optical transformation or computational task are trained/learned through a digital computer using deep learning. Once the training is complete, the layers can be fabricated and assembled to form a 'physical' network that performs the desired computation in a passive manner and at the speed of light propagation. Diffractive networks can achieve universal linear transformations[7–9], and various applications using diffractive processors have been demonstrated such as object classification, pulse processing, imaging through random diffusers, hologram reconstruction, quantitative phase imaging, class-specific imaging, super-resolution image display, all-optical logic operations, beam shaping and orbital angular momentum mode processing, among others[10–30].

While diffractive networks have shown competitive performance on the classification of relatively simpler objects, for example, hand-written digits and fashion products[11], for more complex natural objects such as those from the CIFAR-10 dataset[31], their performance gap compared to the classification accuracy of electronic neural networks is still large[11,32]. Ensemble learning through multiple $D^2$NNs has been demonstrated to improve the inference and generalization of diffractive networks at the cost of reducing the compactness and simplicity of the optical hardware[32].

In this work, we demonstrate, for the first time, a 'time-lapse' image classification scheme with a stand-alone diffractive optical network that significantly enhances the inference and generalization performance of diffractive computing. In this scheme, the objects and/or the diffractive network laterally move relative to each other, either randomly or in a controlled manner, during the detector integration time, enriching the information provided to the diffractive network. In a different context and application, lateral shifts of the object of interest relative to the imager have been routinely used for pixel super-resolution imaging, enhancing the resolution of the reconstructed images[33–36]. Inspired by the success of these pixel super-resolution approaches, here we use the controlled or random relative displacements between the input objects and the diffractive network for time-lapse image classification and report a numerical blind testing accuracy of 62.03% for the classification of grayscale CIFAR-10 images, which constitutes the highest classification accuracy for this dataset achieved so far using a single diffractive optical network. In addition to significantly advancing the inference and generalization performance of $D^2$NNs, these time-lapse diffractive networks can also find broader use in the all-optical processing of spatio-temporal information of a scene or object.

**Results**

The concept of time-lapse image classification with a diffractive network is illustrated in Fig. 1. A diffractive network comprising 5 phase-only diffractive layers, axially separated by $40\lambda$, is placed between the object plane and the detector plane. The detector plane includes 20 detectors[11]: 2 detectors for each class $c$ of the CIFAR-10 dataset, i.e., a 'positive' detector $D_{c,+}$ and a 'negative' detector $D_{c,-}$. The integration time of the output detectors is assumed to be $N\delta_t$, where $N$ is the number of lateral object shifts and each of the $N$ individual shifts has an equal integration time of $\delta_t$. Without changing our conclusions, in alternative implementations, the diffractive network can also laterally move relative to the static object, or both the object and the diffractive network can laterally move at the same time. Each detector $D_{c,\pm}$ is assigned an exponent $\gamma_{c,\pm}$ which operates on the integrated detector power to yield the detector signal $I_{c,\pm}$ (see Fig. 1 and the Methods section). We will report diffractive classification results under two different conditions: (1) the exponents are assumed to be trainable, and (2) non-trainable, fixed as $\gamma_{c,\pm} = 1$. The normalized differential class scores $z_c = \frac{I_{c,+} - I_{c,-}}{I_{c,+} + I_{c,-}}$ are calculated from these detector signals, and the prediction/inference is made in favor of the class receiving the highest differential optical score (see Fig. 1).

For all the D²NNs reported in this work, each trainable diffractive layer consists of 200×200 diffractive elements (diffractive neurons) of size $0.53\lambda \times 0.53\lambda$. The objects are assumed to be phase-only and the diffractive networks are trained using the grayscale CIFAR-10 dataset (refer to the Methods section for details). The hyperparameters that define the grid of lateral displacements of the objects during the time-lapse image classification are $s_{max}$ and $m$, where $s_{max}$ is the maximum (relative) lateral displacement along $x/y$ and $m^2$ refers to the total number of points on the grid, see Fig. 2. The size of the input aperture is another hyperparameter that affects the classification performance of the time-lapse diffractive networks. The impact of these hyperparameters, $s_{max}$, $m$ and the input aperture size, on the performance of time-lapse diffractive classifiers is shown in Fig. 2. The classification performance is quantified by the blind testing accuracy of the networks on 10,000 previously unseen images belonging to the test set of the CIFAR-10 dataset. To obtain each data point in Fig. 2, we trained 3 different diffractive networks with the same hyperparameters and calculated the mean and standard deviation of blind testing accuracies of these 3 trained networks. We see from Fig. 2a that as $s_{max}$ is increased from $3.20\lambda$ to $6.40\lambda$ (while keeping $m = 5$ and the aperture size = $44.8\lambda \times 44.8\lambda$ constant), the mean blind testing accuracy increases until $s_{max} = 5.33\lambda$, where it reaches its highest value of 61.35%. Beyond $s_{max} = 5.33\lambda$, the mean classification accuracy starts to decrease. In Fig. 2b, we set $s_{max} = 5.33\lambda$, aperture size = $44.8\lambda \times 44.8\lambda$ and vary $m$. As $m$ is varied between 3 and 6, the mean accuracy increases rapidly from 58.56% to 61.35% until $m = 5$, beyond which the mean accuracy reaches a plateau. For Fig. 2c, we selected $m = 5$ and $s_{max} = 5.33\lambda$ (as optimized from Figs. 2a-b) and the width of the input aperture was varied between $32.0\lambda$ and $53.3\lambda$. The highest mean accuracy (Fig. 2c) is observed for an input aperture size of $38.4\lambda \times 38.4\lambda$, which is smaller than the object support $44.8\lambda \times 44.8\lambda$. We compared this observation with its counterpart for time-static diffractive image classification (see Supplementary Table S1), where the aperture size corresponding to the highest mean blind testing accuracy is larger than the object support. This comparison indicates that a time-lapse diffractive network prefers a relatively smaller input field-of-view compared to its time-static counterparts.

Next, we juxtapose a time-lapse image classification diffractive network with a time-static diffractive network; see Fig. 3. For this comparison, we chose the time-lapse diffractive network with the best *individual* blind testing accuracy (62.03%) among the networks constituting the results of Fig. 2 and the time-static diffractive network with the best *individual* blind testing accuracy (53.14%) among the networks constituting the results of Supplementary Table S1. For the time-lapse image classification diffractive network, the hyperparameters corresponding to the highest individual accuracy were $m = 5$, $s_{max} = 5.33\lambda$ and input aperture size = $38.4\lambda \times 38.4\lambda$; while for the time-static network, the input aperture size corresponding to best individual accuracy was $51.2\lambda \times 51.2\lambda$. Another difference to be noted between the time-static and the time-lapse diffractive networks chosen for comparison in Fig. 3 is that for the time-static one, the detector exponents were not trainable, i.e., $\gamma_{c,\pm} = 1$, whereas the detector exponents were trainable for the time-lapse network. The reason for this selection is that, unlike the time-lapse diffractive networks, time-static diffractive networks showed overfitting when the detector exponents are trainable, leading to inferior generalization; see Supplementary Table S2.

For an example object from the image class 'ship' (true label: 8), we show in Fig. 3a the detector plane intensity, detector signals and the class scores for the time-static network; similarly, in Fig. 3b we show the time-integral of the detector plane intensity, detector signals and the class scores for the time-lapse image classification network. While the time-static network misclassifies the object for an 'automobile', the time-lapse image classification diffractive network correctly predicts the object to be a 'ship' (predicted label: 8). We also show in Fig. 3c the confusion matrices calculated over 10,000 test images of the CIFAR-10 dataset: the time-lapse image classification diffractive network performs consistently better than the time-static one for all the CIFAR-10 data classes.

Note also that the time-lapse image classification diffractive network designed with *non-trainable* detector exponents (i.e., $\gamma_{c,\pm} = 1$) achieved a blind testing accuracy of 60.35% on the same grayscale CIFAR-10 test dataset (see Supplementary Fig. S1), performing much better than the time-static one for all the CIFAR-10 data classes. The diffractive layers for all these networks are shown in Supplementary Fig. S2.

During the training of the time-lapse diffractive networks, we followed a method similar to the 'dropout' method, which is used in deep learning to reduce overfitting and improve the generalization of a trained model[37]. We defined a hyperparameter $p$ which is the probability that a point on the object-plane grid is 'active' during training, i.e., the probability that the object is positioned at that lateral point during the signal integration at the detector. All the time-lapse networks described thus far were trained with $p = 0.5$. As we describe below, the resilience of the trained time-lapse image classification diffractive networks to deviations from the training settings can be improved by a proper choice of $p$, which is intuitively equivalent to the dropout strategy in deep learning literature.

Related to this hyperparameter $p$, next, we explored the impact of decreasing the number of lateral shifts, $N$, on the blind testing accuracy of time-lapse classifiers: see Fig. 4. The value for each data point in Fig. 4 represents the mean of the classification accuracies over 25 independent blind tests with the same $N$. For Fig. 4a, these $N$ lateral displacements were restricted to coincide with the pre-determined training grid points, and for the case of $N < m^2$, $m^2 - N$ of the $m^2$ lateral shifts were randomly eliminated (not used). For Fig. 4b, however, the $N$ lateral displacements were randomly selected *without following the training grid points*. As we can see in Fig. 4a, the blind testing accuracy decreases as $N$ is decreased; however, the slope of this performance degradation varies depending on the training hyperparameter $p$. For example, in the case of the time-lapse image classification diffractive network shown in Fig. 3b, trained with $p = 0.5$ (green curve in Fig. 4a), the test accuracy drops from 62.03% to 60.69% and 59.37% as $N$ decreases from 25 to 15 and 10, respectively. Compare this with the case of a time-lapse diffractive network trained with $p = 1.0$ (red curve in Fig. 4a), for which the classification accuracy is affected much more severely and decreases from 61.61% to 59.61% and 57.45% as $N$ is decreased from 25 to 15 and 10, respectively. We see that networks trained with lower $p$ values show less sensitivity to decreasing $N$, which is further corroborated by the curves corresponding to two other time-lapse diffractive networks trained with $p = 0.2$ and $p = 0.3$.

Another advantage of training with lower $p$ values is decreased sensitivity to the exact object positions (see Fig. 4b). For Fig. 4b, we selected the $N$ lateral displacements *without* following the training grid points, allowing the object to be displaced (during the time-lapse imaging process) to $N$ arbitrary, randomly selected points within the area $2s_{max} \times 2s_{max}$. In general, for a given $N$, the blind testing accuracies corresponding to such arbitrary displacements (left y-axis of Fig. 4b) are lower than their counterparts for the on-grid displacements shown in Fig. 4a. However, the degradation in classification accuracy, which is shown on the right y-axis of Fig. 4b, is much smaller when $p$ is lower. For example, at $N = 25$, the mean accuracy drop is ~2% for the diffractive network trained with $p = 0.2$, whereas the accuracy drop is ~6% for the $p = 1.0$ diffractive network.

The accuracy of time-lapse diffractive network-based image classifiers for arbitrary lateral displacements of the input objects can be improved by utilizing such random displacements of the objects during the training, rather than training with a pre-determined grid of lateral displacements. For this, the training hyperparameters $p$ and $m$ can be absorbed into a single hyperparameter $N_{tr}$, where $N_{tr}$ refers to the number of arbitrary displacements within $2s_{max} \times 2s_{max}$. To demonstrate this, we trained three time-lapse diffractive networks with $N_{tr} = 10$, $N_{tr} = 15$ and $N_{tr} = 25$ and compared their accuracies for $N = 10$, $N = 15$, and $N = 25$ arbitrary displacements of the input objects, respectively, against the classification accuracies of the time-lapse diffractive networks reported in Fig. 4a-b. The result of this comparison is shown in Fig. 4c: for $N = 10$, $N = 15$ and $N = 25$ arbitrary lateral displacements during the time-lapse imaging process, the mean blind testing accuracies of the corresponding $N_{tr} = N$ diffractive networks are 1.26%, 1.77%, and 1.54%, respectively, higher than the accuracies of the $p = 0.2$ time-lapse diffractive network. This generalization improvement and the inference accuracy increase are due to using arbitrary random lateral displacements of the input objects during the training process instead of blindly applying such random lateral shifts only during the testing phase.

**Discussion**

In previous work, we reported a significant improvement in diffractive network inference performance by ensemble learning and combining the output of several different diffractive networks. For example, mean blind testing accuracies of 61.14% and 62.13% on the CIFAR-10 test set were reported for ensembles of 14 and 30 different D²NNs, respectively[32]. However, the improvement with such a strategy is accompanied by a sacrifice in the compactness of the optical hardware and increased complexity in aligning several diffractive networks within the ensemble. Another shortcoming of ensemble learning of diffractive networks is the large training time. In our previous work, 1252 diffractive models were trained, and ensemble pruning was then performed to arrive at the final design[32]. Time-lapse diffractive network-based image classification provides blind testing accuracies comparable to ensemble learning with only a *single* trained diffractive network. For comparison, the time-lapse diffractive network of Fig. 3b gives 62.03% blind testing accuracy on CIFAR-10 test images. The trade-off for such an advantage is the increase in the imaging/classification time due to the lateral shifts of the objects. However, the alignment and synchronization requirements associated with diffractive network ensembles are evaded. Also, the training of a time-lapse diffractive classifier takes ~20 hours on an NVIDIA GeForce RTX 3090 GPU (see the Methods section), which is orders of magnitude less than the time required to design an ensemble of diffractive networks working together.

Regarding the implementation of time-lapse diffractive network-based image classification, Spatial Light Modulators (SLMs) can be used to perform the lateral displacements of the input objects digitally if a digital representation of each object is available. In an alternative implementation, the diffractive layers and the detectors could be mounted on a movable stage to shift the entire system with respect to the object or input FOV. Perhaps, the simplest implementation of time-lapse diffractive network-based image classification would exploit the natural jitter or movement of the input objects during the integration time of the class detectors. As shown in Fig. 4c, ~60% blind testing accuracy on CIFAR-10 test images can be reached with arbitrary object displacements during the time-lapse inference.

While time-lapse image classification significantly boosts the inference of a single D²NN on the classification of complex objects, there remains plenty of room for improvement to potentially close the large performance gap with their electronic counterparts, convolutional deep neural networks[32]. One possible avenue for such an improvement could be the incorporation of ensemble learning with time-lapse image classification, where the outputs of diversely trained time-lapse D²NNs could be combined for further improvement in generalization and statistical inference. Moreover, in the same way that the time-lapse scheme utilizes the complementary information resulting from the input objects that are laterally shifted, other attributes of light such as polarization or wavelength could also be utilized[9,38]. For example, time-lapse diffractive networks can be trained to work with RGB images instead of grayscale images to benefit from the complementary information carried by different color channels. The incorporation of optical nonlinearities between the diffractive layers of D²NNs could also extend their approximation capability and consequently improve their statistical inference; for further details, see the Supplementary Information of Ref. 6, where the impact of optical nonlinearities within a D²NN architecture was first discussed. All of these constitute possible future directions to explore for further decreasing the performance gap between electronic deep neural networks and D²NNs.

In summary, we reported a time-lapse diffractive network-based image classification scheme for significantly improving the performance of D²NN classifiers with only a single trained diffractive network. The presented time-lapse diffractive network scheme could be vital for realizing compact, low-cost and passive optical processors for all-optical spatio-temporal analysis of information.

**Materials and methods**

**Forward model.** The propagation of coherent light across $K+2$ parallel planes defined by the input (object) plane, $K$ successive diffractive layers, and the output (detector) plane is modeled using the Rayleigh-Sommerfeld theory of scalar diffraction[39], according to which the propagation of a complex wave $U(x, y)$ through a distance $z$ in free-space is described by a linear shift-invariant system with an impulse response defined as follows:

$$h(x, y; z) = \frac{z}{r^2}\left(\frac{1}{2\pi r} + \frac{1}{j\lambda}\right)\exp\left(j\frac{2\pi r}{\lambda}\right)$$

where $\lambda$ is the illumination wavelength, $r = \sqrt{x^2 + y^2 + z^2}$ and $j = \sqrt{-1}$. Upon propagation through the free-space separating layer $l$-1 and layer $l$, the complex field is modulated by the spatially varying complex transmittance $t_l(x, y)$ of layer $l$, i.e.:

$$U(x, y; z_l) = t_l(x, y) \iint h(x - x', y - y'; z_l - z_{l-1}) U(x', y'; z_{l-1}) dx' dy'$$
$$t_l(x, y) = a_l(x, y) \exp(j\varphi_l(x, y))$$

Here, $z_l$ is the axial coordinate of the $l$-th plane, and $l = 1, \cdots, K$, whereas $a_l(x, y)$ and $\varphi_l(x, y)$ are the amplitude and the phase of the complex field transmittance $t_l(x, y)$. For the phase-only diffractive networks reported in this work, $a_l(x, y)$ is assumed to be 1.

In a differential classification scheme, each of the 10 classes of the CIFAR-10 dataset is assigned to two detectors: a virtual positive detector and a virtual negative detector. $D_{c,+}$ ($D_{c,-}$) denotes the active area of the positive (negative) detector assigned to class $c$, $c = 0, 1, \cdots, 9$. For the time-static diffractive networks, the detector signals $I_{c,\pm}$, based on which the class scores are computed, are proportional to the detector powers $P_{c,\pm}$, where

$$P_{c,\pm} = \iint_{D_{c,\pm}} |U(x, y; z_{K+1})|^2 dx dy$$

For the time-lapse diffractive network, light is assumed to be integrated at the detectors over $N$ intervals of $\delta_t$ duration each. The object function ($O$) during the $n$-th interval can be expressed as:

$$O_n(x, y) = O(x - x_n, y - y_n)$$
$$n = 1, \cdots, N$$
$$(x_n, y_n) \in X \times Y$$
$$X = Y = linspace(-s_{max}, s_{max}, m)$$

where $linspace(-s_{max}, s_{max}, m)$ denotes the set of $m$ linearly spaced values between $-s_{max}$ and $s_{max}$. Accordingly, the optoelectronic signals at the detectors are proportional to the integrated photon signals

$$E_{c,\pm} = \alpha \int_0^{N\delta_t} P_{D_{c,\pm}}(t) dt$$

Here, $\alpha$ is an optoelectronic detector-specific constant, and we assume that the propagation delay of light between the object plane and the detector plane is negligible compared to $\delta_t$.

The detectors are assigned the exponents $\gamma_{c,\pm}$, which operate on the optoelectronic signals $E_{c,\pm}$ (after $E_{c,\pm}$ are normalized to have a maximum value of 1) and generate the detector signals $I_{c,\pm}$:

$$I_{c,\pm} = (E_{c,\pm})^{\gamma_{c,\pm}}$$

Finally, the differential class scores are calculated as:

$$z_c = \frac{I_{c,+} - I_{c,-}}{I_{c,+} + I_{c,-}}$$

and the prediction for the object class is defined to be $\arg\max_c z_c$.

**Numerical implementation.** When numerically modeling light propagation through the diffractive networks, the grid spacing along the transverse directions ($x$ and $y$) was chosen to be ~$0.53\lambda$. The Rayleigh-Sommerfeld convolution integrals were computed using the Angular Spectrum Method[39] based on the Fast Fourier Transform (FFT). For all the results presented in this paper, the diffractive networks consisted of 5 phase-only diffractive layers, axially separated by $40\lambda$. Each layer comprised 200×200 diffractive features/neurons, the phases of which were trainable. The (physical) size of each diffractive neuron was assumed to be ~$0.53\lambda \times 0.53\lambda$.

The RGB images in the CIFAR-10 dataset were converted to grayscale to represent the input objects illuminated by a monochromatic and spatially-coherent wave. The objects were resized to span an area of 44.8λ×44.8λ. The object information was assumed to be encoded in the phase channel of the input light, i.e., within the input field of view,

$$U(x, y; z_0) = \exp(j2\pi O(x, y))$$

where $O(x, y)$ is the object function, with its values normalized to lie between 0 and 1. On the output plane, the active area of each detector was assumed to be 6.4λ×6.4λ, and the spacing between the detectors was ~4.27λ along both $x$ and $y$ directions (see Fig. 1).

**Training.** The diffractive networks were trained using the cross-entropy loss function. The differential class-scores $\{z_c\}_{c=0}^{9}$ were converted to probabilities $\{q_c\}_{c=0}^{9}$ over the classes using the softmax function, i.e.,

$$q_c = \frac{\exp(\beta z_c)}{\sum_i \exp(\beta z_i)}$$

where $\beta = 10$ was used. The training loss was defined as:

$$\mathcal{L} = -\sum_{c=0}^{9} \delta_{ck} \log q_c$$

where $k$ is the (true) label, and $\delta_{ck}$ is the Kronecker delta function, i.e., $\delta_{ck} = 1$ if $c = k$ and 0 otherwise.

The trainable parameters of the model were trained by minimizing the loss $\mathcal{L}$ using the Adaptive Momentum ('Adam') stochastic gradient descent algorithm[40]. The forward model was implemented using the open-source deep learning library TensorFlow[41]. The automatic differentiation functionality of TensorFlow was exploited to facilitate the gradient computations for optimization. A batch size of 8 was used to implement the stochastic gradient descent. The built-in TensorFlow implementation of Adam optimizer was used with the default values except for the learning rate, which had an initial value of 0.001 and was reduced by a factor of 0.7 every 8 epochs.

All the networks were trained for 100 epochs using 45000 images from the training set of the CIFAR-10 dataset. The remaining 5000 images of the CIFAR-10 training set were left out for validation, i.e., after every epoch, the accuracy of the model on these 5000 images was evaluated. The model state at the end of the epoch for which the validation accuracy was maximum was ultimately used for blind testing.

The training time of the time-lapse diffractive networks depended upon the hyperparameters $m$ and $p$. For $m = 5$ and $p = 0.5$, the training took ~20 hours on an NVIDIA GeForce RTX 3090 GPU in a machine running on Windows 10.

**Figures**

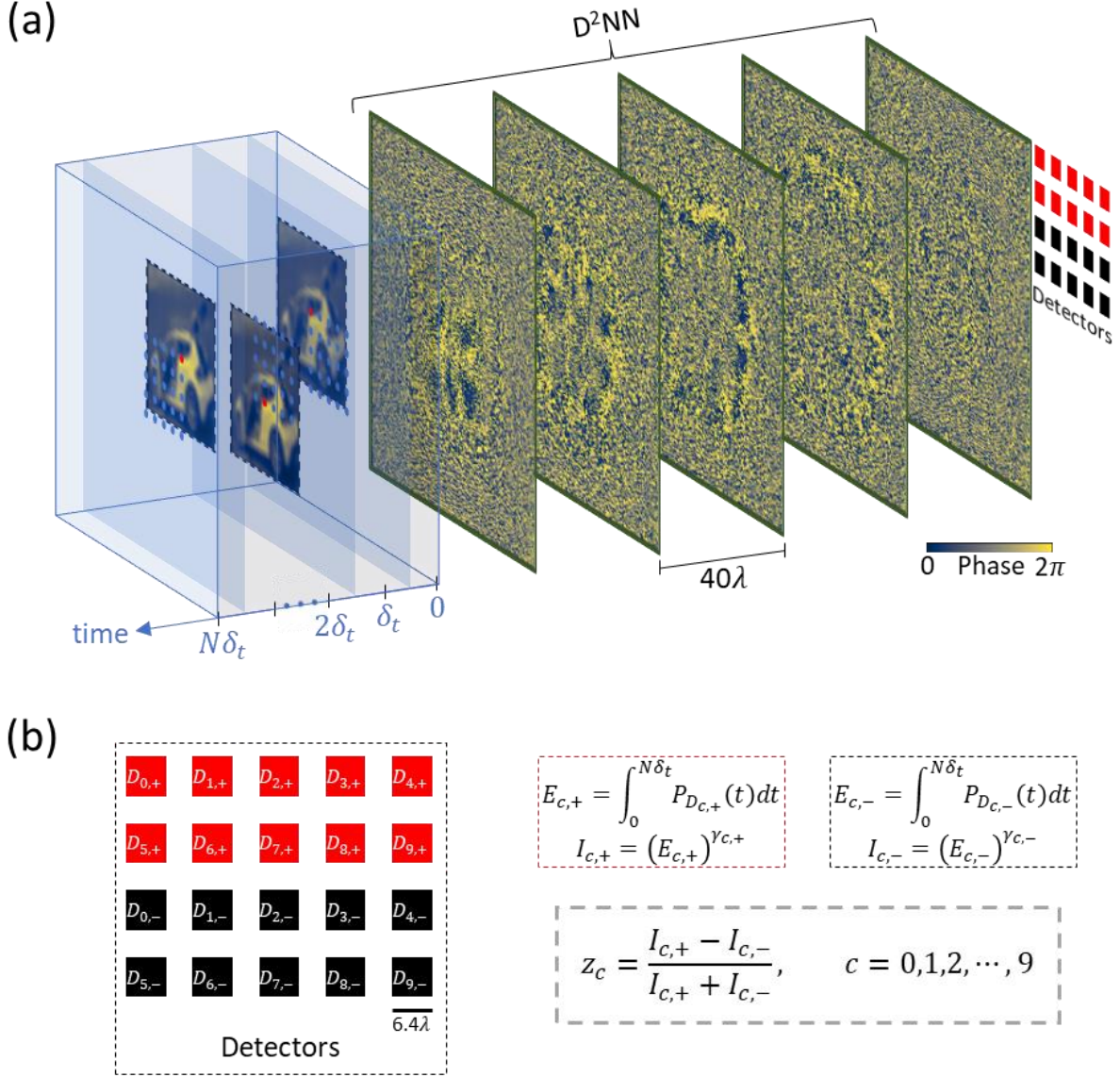

**Fig. 1 Time-lapse image classification using a D²NN.** (a) A diffractive network with 5 phase-only diffractive layers followed by 20 detectors at the detector plane for differential image classification. The integration time of each detector is $N\delta_t$. During each one of the $N$ intervals of $\delta_t$ duration, the center of the object is laterally displaced to a new point (red circle); these lateral displacements can be entirely random or follow a predefined grid (blue circles). (b) Labeling of the detectors where $D_{c,+}$ ($D_{c,-}$) denote the positive (negative) detectors assigned to class $c$. The differential class-scores $z_c$ are used for the final classification decision based on the maximum score.

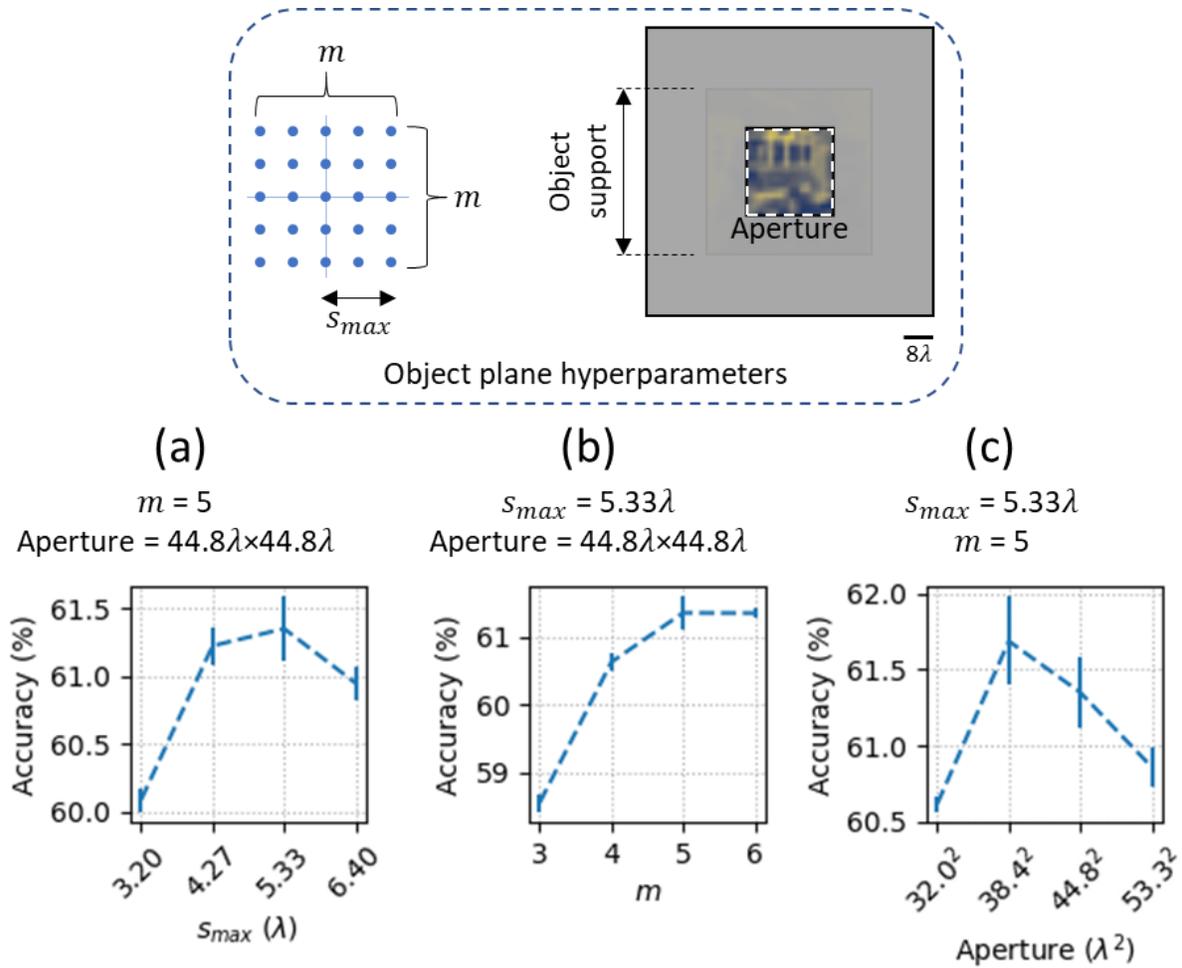

**Fig. 2 Dependence of the time-lapse image classification D²NN performance on the hyperparameters of the input plane.** Top left: Grid of points representing the lateral displacements of the input objects during the training, defined by the two hyperparameters, $s_{max}$ and $m$. $s_{max}$ represents the maximum displacement along either the vertical or the horizontal direction, whereas $m^2$ is the total number of grid points. Top right: the dashed white square represents the input aperture immediately following the object, the area of which is another hyperparameter. (a) Dependence of the blind testing accuracy on $s_{max}$ with $m$ and the input aperture kept constant. (b) Effect of $m$ on the blind testing accuracy of the trained time-lapse diffractive classifiers as $s_{max}$ and the input aperture are kept constant. (c) Dependence of the blind testing accuracy on the input aperture size while $s_{max}$ and $m$ are kept constant. For (a)-(c), the data points and the error bars represent the mean and the standard deviation values, respectively, calculated from three designs, which are obtained by training three different time-lapse D²NN classifiers for the same set of hyperparameter values. The curves are linearly interpolated between the data points.

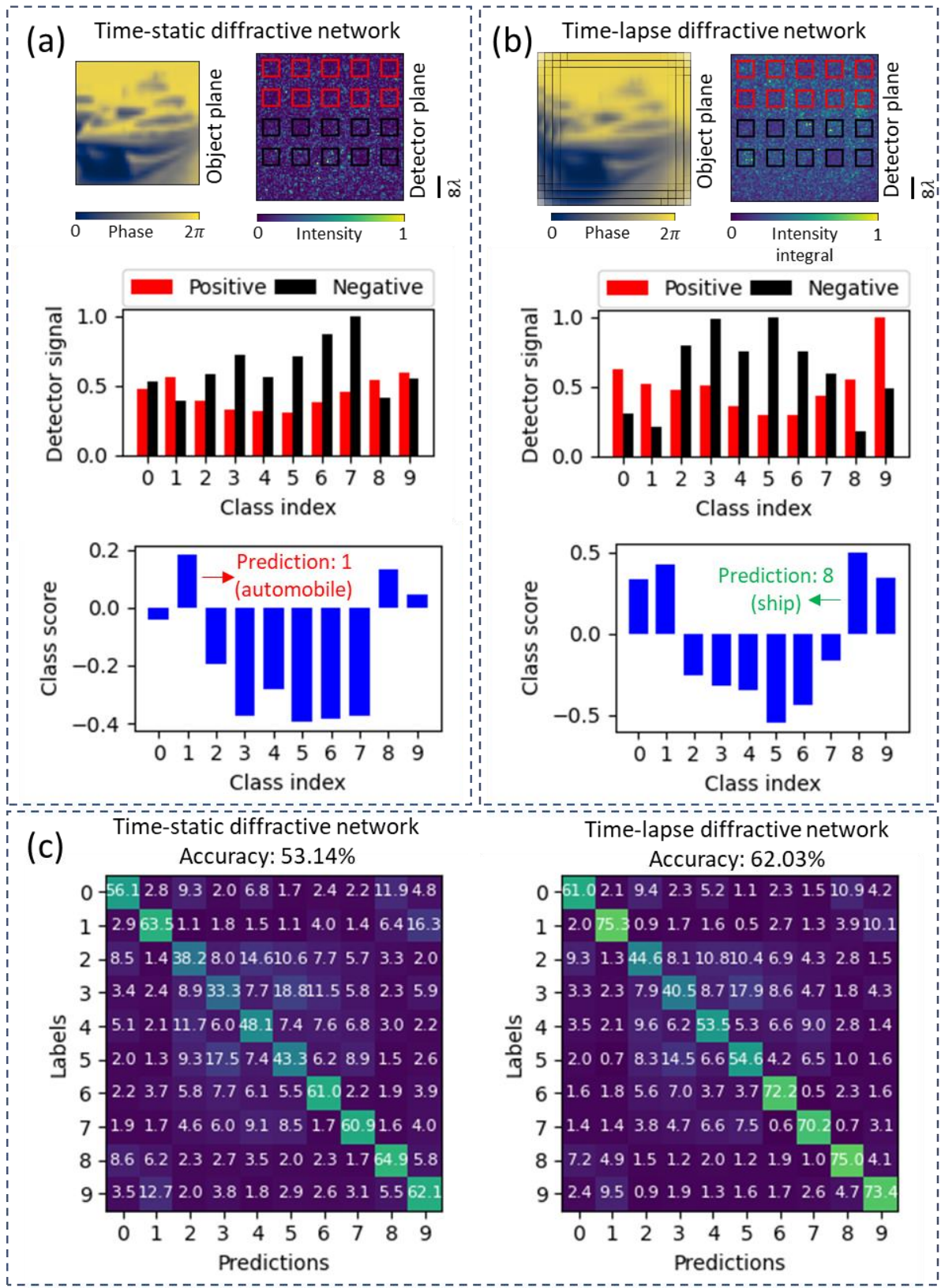

**Fig. 3 Comparison between a time-static and a time-lapse diffractive image classifier.** (a) Detector

plane intensity, detector signals, and class scores from the time-static network for an object (a ship) from the data class 8 of the CIFAR-10 dataset. Based on the class scores, the object is misclassified to be an automobile. (b) Integrated detector plane intensity, detector signals, and class scores from the time-lapse diffractive network for the same object, which is correctly classified as a ship. (c) Confusion matrices for the two networks, evaluated by blind testing on 10000 CIFAR-10 test images. The overall accuracies of the networks are 53.14% and 62.03%, respectively. The training parameters of these two networks are shown in Supplementary Fig. S2a-b.

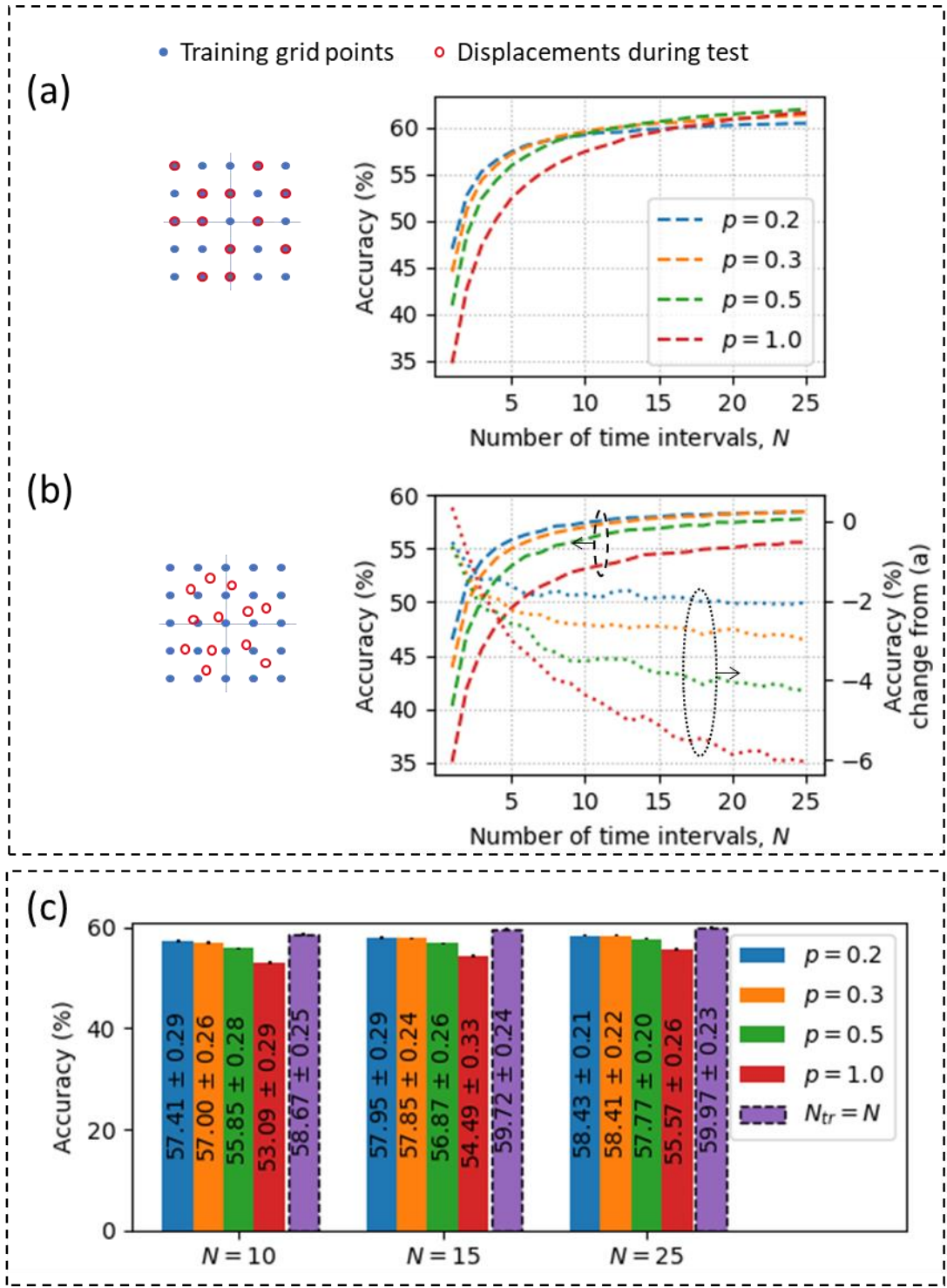

**Fig. 4 Impact of the lateral shifts.** (a) Impact of decreasing the number of lateral shifts ($N$) on the blind testing accuracy while confining the input object displacements to the training grid points; the time-lapse D$^2$NNs were trained with different $p$ values. Diffractive networks trained with lower $p$ values are less affected by a decrease in $N$. (b) Impact of *random off-grid* lateral displacements of input objects during the blind testing of time-lapse diffractive networks trained with displacements confined to a predefined grid. Time-lapse diffractive classifiers trained with lower $p$ values are more resilient to such deviation

from the training settings. (c) Improvement of blind testing accuracies for a given $N$, by training the time-lapse diffractive network with $N_{tr} = N$ arbitrary/random displacements within the range $2s_{max} \times 2s_{max}$ instead of training with a set of fixed lateral displacements defined by a pre-determined lateral grid. For (a)-(c), the values (errors) corresponding to the data points represent the mean (standard deviation) values calculated through the blind testing of the same trained network 25 times, every time with $N$ arbitrary lateral displacements of the input objects.